\documentclass[preprint,nofootinbib,superscriptaddress,amssymb]{revtex4}
\usepackage{amsmath} 
\usepackage{verbatim} 
\usepackage{graphicx} 
\usepackage{subfig} 
\usepackage{float} 

\begin{document}

\title{Structure Formation Constraints on \\ Sommerfeld-Enhanced Dark Matter Annihilation}

\author{Cristian Armendariz-Picon}
\affiliation{Department of Physics, Syracuse University, Syracuse, NY 13244-1130, USA}

\author{Jayanth T. Neelakanta}
 \affiliation{Department of Physics, Syracuse University, Syracuse, NY 13244-1130, USA}
 
\begin{abstract}
We study the growth of cosmic structure under the assumption that dark matter self-annihilates with an averaged cross section times relative velocity that grows with the scale factor, an increase known as Sommerfeld-enhancement.  Such an evolution is expected in  models in which a light force carrier in the dark sector enhances the annihilation cross section of dark matter particles, and has been invoked, for instance, to explain  anomalies in cosmic ray spectra reported in the past. In order to make our results as general as possible, we assume that dark matter annihilates into a relativistic species that only interacts gravitationally with the standard model. This assumption also allows us to test whether the additional relativistic species mildly favored by cosmic-microwave background data  could originate from dark matter annihilation. We do not find evidence for Sommerfeld-enhanced dark matter annihilation and derive the corresponding upper limits on the annihilation cross-section. 

\end{abstract}

\maketitle

\section{Introduction}

Although there are many different and well-motivated  dark matter models, current observational constraints on dark matter properties have not pinned down the  actual microscopic origin of dark matter yet. Indeed, from a purely phenomenological perspective, a vast array of cosmological and astrophysical observations can be accommodated by a simple model in which dark matter is modeled as a non-interacting pressureless fluid of unknown origin. 

The simplest way to  explain the properties of such a fluid is to assume that dark matter consists of non-interacting and non-relativistic particles. In this scenario, the amount of dark matter in our universe is a free parameter that has to be chosen to fit observations and thus remains unexplained. On the other hand, if dark matter particles are assumed to self-annihilate with an averaged cross section times relative velocity of the order of the weak scale, 
\begin{equation}\label{eq:sigma0}
\langle \sigma v\rangle_w \equiv 
	3\cdot 10^{-26} \mathrm{cm}^3\, \mathrm{s}^{-1},
\end{equation}
dark matter particles decouple from radiation in the early universe while being non-relativistic, with an abundance that roughly fits the observed amount of dark matter,
\begin{equation}\label{eq:abundance}
	\Omega_c h^2\approx 0.1 \frac{\langle \sigma v\rangle_w}{\langle \sigma v\rangle}.
\end{equation}
This equation holds  regardless of the precise value of the dark matter mass and the particles dark matter annihilates into. In this scenario, we not only explain the major properties of dark matter, but also its amount. This is why weakly interacting massive particles (wimps) are widely believed to be the dark matter constituents.

But somewhat recently, motivated by certain anomalies in cosmic ray spectra \cite{Adriani:2008zr,Chang:2008aa,Abdo:2009zk}, several authors have suggested that the dark matter self-annihilation rate today may  differ from the rate suggested by equation (\ref{eq:abundance}) \cite{Cirelli:2008pk,MarchRussell:2008tu,ArkaniHamed:2008qn}. If $f$ is the fraction of the energy deposited into standard model particles by  two annihilating dark matter wimps, these models require \cite{Cirelli:2008pk, Cholis:2008hb}
\begin{equation}
 f\cdot \langle \sigma v\rangle \sim  10^2 \, \langle \sigma v \rangle_w
\end{equation}
for a wimp of mass $m\sim1 \, \mathrm{TeV}$. Therefore, in order to preserve the successful postdiction of the dark matter abundance, these authors have suggested that  the dark matter annihilation rate is inversely proportional to the dark matter velocity, and thus increases as the universe expands and the velocity redshifts. 
\begin{equation}
\langle \sigma v \rangle \propto \frac{1}{v}.
\end{equation}
 A simple way to accomplish such an increase involves the  Sommerfeld enhancement of the annihilation cross section induced by a new, sufficiently light force carrier \cite{Sommerfeld, Hisano:2004ds}. 

Recombination  places quite stringent constraints on the annihilation cross section of the enhanced dark matter models. If dark matter efficiently annihilates into radiation during recombination,  the injection of this radiation into the plasma significantly affects the temperature anisotropies in the cosmic microwave background radiation. Using this effect, several groups  have been able to place an upper limit on the thermally averaged annihilation rate times velocity during recombination \cite{Slatyer:2009yq,Galli:2011rz},
\begin{equation}\label{eq:rec limit}
	\langle \sigma v \rangle \leq 120
	\frac{ \langle \sigma v\rangle_w}{f} \, 
	\frac{m c^2}{\mathrm{TeV}}
	\quad  \text{at} \quad 95\% CL.
	\end{equation}
On the face of this limit, models that explain cosmic ray 	anomalies with enhanced annihilation cross section are already ruled out or on the verge of being ruled out by  forthcoming PLANCK data \cite{Ade:2011ah}. 

Unfortunately, the limit on the annihilation cross section (\ref{eq:rec limit})  depends on the model-dependent parameter $f$, which can vary by several orders of magnitude. In those (nearly ruled out) models that attempt to explain the cosmic ray anomalies mentioned above, $f$ is of order one, whereas in models in which  dark matter is part of a dark sector that  interacts only gravitationally with the standard model, $f$ vanishes. In  extreme cases like the latter, the limit (\ref{eq:rec limit}) is not very useful.

In this article we set limits on the dark matter annihilation cross section that do not depend on $f$, and thus apply to a wider class of dark matter candidates, beyond those designed to address the aforementioned cosmic ray  anomalies. Our constraints are based on the impact of dark matter annihilation on the formation and growth of large-scale structure, including the cosmic microwave background anisotropies and the distribution of dark matter. Because the presence of additional force carriers in the dark sector still remains well-motivated, regardless of the dark matter annihilation channels, and because models with enhanced annihilation cross section  provide distinct phenomenological signatures we  focus on dark matter that self-annihilates into  dark radiation with a Sommerfeld-enhanced  cross section (several  specific models in this class have been studied for instance in  \cite{Ackerman:2008gi, Blennow:2012de}.) Our dark radiation is assumed to not interact with standard model particles, which corresponds to the limit $f=0$ in the class of models discussed above. Hence, any imprint of annihilation on cosmic observables must come from either the suppressed growth of dark matter structures, or from the gravitational interactions of its annihilation products, which are  present in any scenario in which dark matter self-annihilates.  

For negliglible values of $f$, one can  also derive  quite stringent constraints on the self-scattering cross-section of dark matter (which should also experience Sommerfeld enhacement), because the latter would cause the central cores of gravitational bound astrophysical systems to become spherical, rather than elliptical, in conflict with observations \cite{Buckley:2009in, Feng:2009hw}. Unfortunately however, there is no model-independent relation between the scattering and annihilation cross sections, so these constraints cannot be directly applied to self-annihilation. In addition, these constraints only limit the scattering cross section at velocities  of the order found in the corresponding dark matter halo. In contrast, our limits on annihilation do not depend on the dark matter velocity, and only rely on the assumption that dark matter is non-relativistic. 

Our considerations of dark matter annihilation with a Sommerfeld-enhanced cross section are further motivated by two seemingly unrelated phenomenological problems. On one hand, it has been argued for some time that  in the standard $\Lambda$CDM model the central densities of dark matter haloes, and the number of small  subhaloes, do not appear to match observations \cite{Moore:1994yx,Klypin:1999uc}, although this eventual disagreement may have conventional astrophysical explanations  \cite{Mashchenko:2006dm,Bullock:2000wn}. A natural way to explain the  discrepancy is to assume that dark matter interacts or annihilates with a cross section that is inversely proportional to the dark matter velocity, as in Sommerfeld-enhanced models \cite{Yoshida:2000uw, Aarssen:2012fx}, or simply to assume that dark matter self-annihilates with  cross section larger than that required by equation (\ref{eq:abundance})  \cite{Kaplinghat:2000vt}. On the other hand, it has also been noticed that cosmic microwave data seem to indicate an additional  relativistic dark component that  interacts only gravitationally with the standard model (see for instance \cite{Komatsu:2010fb,Hamann:2010bk, Dunkley:2010ge, Archidiacono:2011gq}). It is thus worthwhile to investigate whether this additional radiation could originate from dark matter annihilation, a circumstance that would link these two apparently unrelated problems.

In the context of the original Sommerfeld-enhancement  models designed to explain the cosmic ray anomalies, our limits can be used for instance to determine the values of $f$ for which the effects of dark matter annihilation on structure formation have to be taken into account.  In the general case, they help   further constrain the properties of the yet to be identified dark matter particle, and, eventually, may explain the origin of the dark radiation hinted at by  cosmic microwave data.

\section{Annihilating Dark Matter}

As we mentioned in the introduction, for our purposes dark matter is well described by a  pressureless perfect fluid, with energy momentum tensor
\begin{equation}
	T^{(c)}_{\mu\nu}=\rho_c u_\mu^{(c)} u_\nu^{(c)},
\end{equation}
where $\rho_c$ is the energy density of dark matter, and $u_{(c)}^\mu$ its four-velocity, $g^{\mu\nu}u_\mu^{(c)} u_\nu^{(c)}=-1$. By assumption, the pressure of dark matter vanishes.  In appendix \ref{sec:Boltzmann Equation} we link this perfect fluid description to a kinetic description, in which dark matter is regarded as an ensemble of non-relativistic particles.  Our goal is to study the effects of dark matter annihilation on the growth of structure.  For simplicity, we assume that dark matter annihilates into relativistic particles that  interact only gravitationally with the standard model, but interact sufficiently rapidly with other particles in the dark sector (or themselves) to justify a perfect fluid approximation on the scales of interest. In that sense, the behavior of dark radiation mimics the behavior of  photons prior to recombination. We shall thus regard  the dark matter annihilation products as a perfect relativistic dark fluid, with energy-momentum tensor
\begin{equation}
	T^{(d)}_{\mu\nu}=(\rho_d+p_d) u_\mu^{(d)} u_\nu^{(d)}+ p_d\,  g_{\mu\nu}, \quad \text{where}
	\quad p_d=\frac{\rho_d}{3}.
\end{equation}
As it turns out, present cosmic microwave anisotropy data  suggest the existence of such an additional relativistic species (see for instance \cite{Archidiacono:2011gq}). Our dark radiation provides a natural candidate for this additional relativistic component for three reasons: $i)$~Since dark matter is negligible during early radiation domination, its annihilation products are unlikely to conflict with the successful predictions of big-bang nucleosynthesis. $ii)$~Cosmic microwave anisotropy data probe times during which the amount of dark matter was sizable. $iii)$~As we shall see, with a Sommerfeld-enhanced annihilation cross section, dark matter does not entirely freeze out at early times, but keeps annihilating until the dark-matter dominated era.  Note that  studies suggesting the presence of an additional dark relativistic species  typically model this radiation as collisionless (neutrino-like) \cite{Komatsu:2010fb,Hamann:2010bk, Dunkley:2010ge, Archidiacono:2011gq}. Although in this case  a hydrodynamical description breaks down at  small scales, this difference in description should not have much of an impact on cosmological observables, because dark radiation is not visible and never becomes the dominant component of the universe.

In the absence of particle number violating interactions, the energy-momentum tensor of dark matter is covariantly conserved, but in the presence of annihilation, dark matter particles transfer energy to its annihilation products. To determine the energy  lost by the dark matter fluid, we rely on the kinetic description of appendix \ref{sec:Boltzmann Equation}, which yields 
\begin{subequations} 
\begin{equation}\label{eq:c conservation}
	\nabla_\nu T^{(c)}_\mu{}^\nu=-\frac{\langle \sigma v\rangle }{m} \rho^2_c u^{(c)}_\mu.
\end{equation}
Here, $\langle \sigma v\rangle$ is the average dark matter annihilation cross section times relative velocity defined in equation (\ref{eq:sigma v}), and $m$ is the dark matter particle mass.
Note that the rates at which energy and momentum are lost are inversely proportional to $m$, because the annihilation rate is proportional to the square of the number density, and the energy density is proportional to the mass $m$. The energy lost by the dark matter fluid due to annihilation is  gained by the dark radiation fluid, so
\begin{equation}\label{eq:d conservation}
	\nabla_\nu T^{(d)}_\mu{}^\nu=+\frac{\langle \sigma v\rangle }{m} \rho^2_c u^{(c)}_\mu.
\end{equation}
\end{subequations}
In this way, the combined energy-momentum tensor of cold dark matter and dark radiation remains covariantly conserved.

\subsection{Background Evolution}
\label{sec:Background Evolution}

We turn our attention now to the evolution of the dark matter and dark radiation energy densities in an unperturbed, spatially flat FRW universe,
\begin{equation}\label{eq:unperturbed FRW}
	ds^2=a^2(\tau)\left[-d\tau^2+d\vec{x}\,^2\right].
\end{equation}
The equations of motion  of dark matter and dark radiation are given by the time components of equations (\ref{eq:c conservation}) and (\ref{eq:d conservation}), with the four velocities of dark matter and dark radiation taken to be $u^\mu=\delta^\mu{}_0/a$,
\begin{subequations}\label{eq:back evolution}
\begin{align}
\rho_c'+3\mathcal{H}\rho_c &=-\frac{\langle \sigma v\rangle}{m}\rho_c^2 a, \label{eq:dm evolution}\\
\rho_d'+4\mathcal{H}\rho_d &=+\frac{\langle \sigma v\rangle}{m}\rho_c^2 a.
\end{align}
\end{subequations}
We have defined $\mathcal{H}=a'/a$, and a prime denotes a derivative with respect to conformal time $\tau$. In a spatially flat universe we are free to choose the value of $a$ today, which we set to one. Equations (\ref{eq:back evolution}) hold after the  kinetic decoupling of dark matter, which typically occurs well before nucleosynthesis \cite{Dent:2009bv,Feng:2010zp}.

If the evolution of the scale factor is known,  equations (\ref{eq:back evolution}) can be readily integrated to give the evolution of dark matter and dark radiation,
\begin{subequations}\label{eq:bkgd evolution}
\begin{align}
\label{eq:c bkgd evolution}
\rho_c &=\frac{\rho_c^i a_i^3}{a^3}\left(1+\rho_c^i a_i^3\int_{\tau_i}^\tau d\tau'
\frac{\langle \sigma v\rangle}{m} \frac{1}{a^2}\right)^{-1} \\
\label{eq:d bkgd evolution}
\rho_d & =\frac{1}{a^4}\int_{\tau_i}^\tau d\tau'\,  
\frac{\langle \sigma v\rangle}{m}\frac{(\rho_c^i a_i^3)^2}{a} 
\left(1+\rho_c^i a_i^3\int_{\tau_i}^{\tau'} d\tau''
\frac{\langle \sigma v\rangle}{m} \frac{1}{a^2}\right)^{-2},
\end{align}
\end{subequations}
where $\rho_c^i$, $a_i$ and $\tau_i$ are integration constants, and we have assumed that sufficiently early, at $\tau=\tau_i$, the amount of dark radiation is negligible, $\rho_d^i=0$. With this choice, dark radiation only originates from dark matter annihilation; a non-zero value of $\rho_d^i$ would lead to an additional contribution to the dark radiation density that may or may not have originated from the former.  

In this article, we mostly concentrate on the regime in which Sommerfeld enhancement operates, when, according to the discussion in  appendix \ref{sec:Boltzmann Equation}, the averaged relative velocity between dark matter particles $v_\mathrm{rel}$ lies in the appropriate interval next to equation (\ref{eq:S enhancement}). We therefore assume that dark matter annihilates into dark radiation with a Sommerfeld-enhanced cross section, which, according to equation (\ref{eq:SE sigma v}), is proportional  to the scale factor,
\begin{equation}\label{eq:sigma a}
\frac{\langle \sigma v\rangle}{m}=\Gamma a,
\end{equation}
with constant $\Gamma$. 
Although the times at which Sommerfeld enhancement operates are strongly model-dependent, we note that  dark matter particles typically decouple from the thermal bath well before big-bang nucleosynthesis, so we expect their velocities to be  below the velocity $v_0$ introduced in the appendix, certainly by  nucleosynthesis. In our numerical solutions, we therefore assume that Sommerfeld enhancement is already operating at an initial scale factor $a_i=10^{-10}$.

Clearly, in the presence of annihilation, the density of dark matter decreases faster than it otherwise would.  For a cross section of the form (\ref{eq:sigma a}), the density of dark matter during radiation domination is, for instance,
\begin{equation}\label{eq:rho c rad dom}
\rho_c=\rho_c^i\left(\frac{a_i}{a}\right)^3
\left(1+\frac{\langle \sigma v\rangle_i}{m}\frac{\rho_c^i}
{H_i}\log\frac{a}{a_i}\right)^{-1},
\end{equation}
where, again, the subindex $i$ denotes the initial value of the corresponding quantity. In contrast to what happens in the conventional freeze-out scenarios,  the correction factor proportional to $\langle \sigma v\rangle_i$  slowly varies for $a\gg a_i$, suggesting that annihilation keeps operating during radiation domination. Also note that the dark matter density diverges at a scale factor $a<a_i$.  Of course,  at early times our description of cold dark matter ceases to valid, because at sufficiently high densities we are not supposed to ignore inverse annihilations and other processes responsible for keeping the dark matter density in local thermal equilibrium.

To proceed with our analysis, we assume that the annihilation cross section is sufficiently small. On general grounds, we expect the quantitative effects of annihilation to be controlled by the relative change in the number of particles in a comoving volume during a Hubble time,
\begin{equation}\label{eq:impact}
 R\equiv 
 -\frac{1}{\mathcal{H}}\frac{d\log (a^3 \rho_c)}{d\tau}=
	\frac{\langle\sigma v\rangle}{m} \frac{\rho_c}{H}.
\end{equation}
This is, for instance, the case in equation (\ref{eq:rho c rad dom}), in which this factor appears explicitly in the correction to the energy density. Therefore, $\langle \sigma v\rangle$ is small if $R$ remains much smaller than one throughout cosmic history. In that case, it is enough to calculate the impact of annihilation on any cosmological variable just to first order in $\langle \sigma v\rangle$.  Note that to leading order in $\langle \sigma v\rangle$, $R$ is constant during radiation domination, and proportional to $a^{-1/2}$ during matter domination.

To see how this works, consider for instance the amount of dark radiation. Neglecting the higher order correction in the denominator of the integrand in (\ref{eq:d bkgd evolution}) we find 
\begin{equation}\label{eq:attractor}
\rho_d\approx \frac{\langle \sigma v\rangle}{m}\rho_c^2 a \cdot(\tau-\tau_i).
\end{equation}
This equation shows that in this limit the amount of dark radiation does not depend on $\tau_i$ for $\tau\gg \tau_i$, and that $\rho_d$ actually scales  like non-relativistic matter instead of  radiation. In the same limit, the fraction of the total radiation in the dark form during radiation domination is
\begin{equation}
	\frac{\rho_d}{\rho_r}\approx 
	\frac{\langle\sigma v\rangle}{m} 
	\frac{\rho_c}{H} \, 
	\frac{\Omega_c}{1-\Omega_c},
\end{equation}
showing that for an $R$ of order one, the amount of dark radiation is negligible during big-bang nucleosynthesis, but  becomes sizable, about $10\%$ at redshifts of about $z\approx 5z_\mathrm{eq}$, where $z_\mathrm{eq}$ is the redshift of matter-radiation equality. This is relevant because scales entering the horizon at that time are probed by cosmic microwave temperature multipoles of about $\ell\approx 700$, which roughly corresponds to the region probed by WMAP cosmic microsave anisotropy data \cite{Komatsu:2010fb}.

Equations (\ref{eq:bkgd evolution}) are useful during radiation domination, when the scale factor is explicitly known. In order to determine how the energy density of dark matter evolves during matter domination, we  introduce the scale factor $a$ as a time variable in equation (\ref{eq:dm evolution}). To integrate the resulting expression we use Friedmann's equation, neglecting both  standard and dark radiation. The solution is
\begin{equation}\label{eq:dm mat dom}
	\rho_c\approx \rho_c^i \left(\frac{a_i}{a}\right)^3
	\left[1+\frac{\langle \sigma v\rangle_i}{m}
	\frac{\rho_c^i}{H_i}
	\left(1-\frac{a_i^{1/2}}{a^{1/2}}\right)\right]^{-2},
\end{equation}
where a subindex $i$ denotes the value of the corresponding quantity at an arbitrary scale factor $a_i$. Therefore, as opposed to what happens during radiation domination,  dark matter freezes out at $a\gg a_i$, when  its density decays as in the absence of annihilation.   From equation (\ref{eq:attractor}), the amount of dark radiation is simply
\begin{equation}
\rho_d=2\frac{\langle \sigma v\rangle}{m}\frac{\rho_c^2}{H}.
\end{equation}

The time of matter-radiation equality   depends on $\langle\sigma v\rangle$, because both the amount of dark matter and dark radiation depend on the latter. Since  $R$ is proportional to $a^{-1/2}$ during matter domination, we do not expect the values of $\langle \sigma v\rangle$ long after matter-radiation equality to significantly affect cosmological observables, even under the assumption that $\langle \sigma v\rangle$ has been growing with the scale factor since that time.  This is important because, as we discuss in  appendix \ref{sec:Boltzmann Equation}, $\langle \sigma v\rangle$ should become constant at late times, presumably during matter domination. We do not incorporate this saturation in our model, however, so as to avoid an excessive proliferation of free parameters. 

To conclude our analysis of the background evolution, let us consider the effect of annihilation on the age of the universe,
\begin{equation}
t_0=\frac{1}{H_0} \int_0^1 \frac{da}{a\sqrt{\Omega_\Lambda^0+\Omega_b^0 a^{-3}+ \rho_c(a)/\rho_\mathrm{crit}^0 +\Omega_r^0 a^{-4}+\rho_d(a)/\rho_\mathrm{crit}^0}},
\end{equation}
where $\rho_\mathrm{crit}^0$ is the critical density today. Clearly, for fixed values of the remaining cosmological parameters (including the dark matter density today), an increase in $\Gamma$ causes an increase in $\rho_d$, and also induces an increase in $\rho_c$ at earlier times. Therefore, such a change lowers the age of the universe.

\subsection {Linear Perturbations}

Our main concern here is the impact of annihilating dark matter on the formation of structure in the linear regime. We thus consider linear perturbations around the FRW spacetime (\ref{eq:unperturbed FRW}), and decompose  them in Fourier modes,
\begin{equation}\label{eq:Friedmann perturbed}
	ds^2=a^2\left[-d\tau^2+(\delta_{ij}+h_{ij})dx^i dx^j\right],
	\quad
	h_{ij}=\frac{k_i k_j}{k^2} h+6 \left(\frac{k_i k_j}{k^2}-\frac{1}{3} \delta_{ij}\right)\eta.
\end{equation}
Here, $\eta$ and $h$ are the conventional metric potentials in synchronous gauge, which we adopt to connect  our equations with the numerical results presented below. 

Because the energy momentum tensors of dark matter and dark radiation still have perfect fluid form, the linearized Einstein equations retain their conventional form, and we shall not write them down here (see for instance \cite{Ma:1995ey} for the explicit equations.) 
We shall primarily address the modifications that annihilation imposes on the dynamics of both dark matter and radiation. The linearized time and spatial components of equation (\ref{eq:c conservation}) for dark matter are
\begin{subequations}
\begin{align}\label{eq:delta c prime}
	\delta_c'+\frac{1}{2}h'-k^2 v_c
	+\frac{\delta\langle \sigma v\rangle}{m}\rho_c\, a
	+\frac{\langle \sigma v\rangle}{m}\rho_c \delta_c\, a &=0,\\
	\label{eq:vc}
	v_c'+\mathcal{H}v_c&=0,
\end{align}
\end{subequations}
where $\delta\langle \sigma v\rangle$ is given by equation (\ref{eq:delta sigma v}),  and we define  velocity potentials by $u_i\equiv a\, \partial_i v$ where $u_i$ are the spatial components of the four-velocity.  Note that the annihilation cross section does not enter the equation for the velocity perturbation, which admits 
\begin{equation}\label{eq:vc zero}
v_c=0
\end{equation}
as a solution.  Therefore, as in the absence of annihilation, we can use the residual gauge freedom of synchronous gauge to set $v_c=0$. In this gauge, the  equations of motion for dark radiation simplify to
\begin{subequations}
\begin{align}
\delta_d'+\frac{2}{3}h'-\frac{4}{3}k^2 v_d
-\frac{\delta\langle \sigma v\rangle}{m}\frac{\rho_c^2}{\rho_d}\,a
-\frac{\langle \sigma v\rangle}{m} \frac{\rho_c^2}{\rho_d}(2\delta_c-\delta_d) \, a &=0,\\
\label{eq:vd prime}
v_d'+\frac{1}{4}\delta_d+\frac{\langle \sigma v\rangle}{m}\frac{\rho_c^2}{\rho_d}v_d\,  a&=0.
\end{align}
\end{subequations}
Note that if $\langle \sigma v\rangle$ is time-dependent, it is not consistent to assume that its fluctuations $\delta\langle \sigma v\rangle$ vanish. Indeed, as we argue in  appendix \ref{sec:Boltzmann Equation}, in the non-relativistic limit we should set, to leading order in couplings,
\begin{equation}\label{eq:delta sigma v}
	\delta\langle \sigma v\rangle=\langle \sigma v\rangle \, \frac{h}{6}.
\end{equation}
Heuristically, with $\langle \sigma v\rangle=m \Gamma a$, a perturbation in the scale factor $a\to a+\delta a$ induces a perturbation in the averaged cross section $\delta\langle \sigma v\rangle=\langle \sigma v \rangle \delta a/a$. But on large scales (in cosmic time coordinates) such a perturbation is equivalent to a metric perturbation with $h=6\,\delta a/a$, from which equation (\ref{eq:delta sigma v}) automatically follows.

\subsection{Initial Conditions}
\label{sec:Initial Conditions}
In order to calculate the impact of dark matter annihilation on the  temperature anisotropies and the distribution of matter, we need to specify initial conditions for  the perturbations in all the components of the universe, including dark mater and dark radiation. These initial conditions are set well into the radiation-dominated era, when all modes of cosmological interest are much larger than the Hubble radius. 

At present, the angular correlations of cosmic microwave background temperature anisotropies are well-fit by a nearly scale-invariant spectrum of \emph{adiabatic} primordial perturbations, in agreement with the predictions of the arguably simplest (single field) inflationary models. We would therefore like to impose adiabatic initial conditions on our perturbations, which we expect to be different for  dark matter and dark radiation. 

It turns out that in the presence of annihilation,  and in synchronous gauge, the question of adiabaticity is a subtle one. Weinberg has shown for instance that the linearized perturbation equations in longitudinal gauge always admit (under rather mild assumptions) an ``adiabatic" solution in the long wavelength limit $k\to 0$ \cite{Weinberg:2003sw}. The form of this adiabatic solution is explicitly known, regardless of the dynamics of the universe constituents, and this makes it straightforward to impose adiabatic initial conditions in longitudinal gauge, even in the presence of annihilation. But if one transforms this longitudinal adiabatic solution to synchronous gauge one finds that the total energy density perturbation vanishes, while $\eta$ remains finite. Although this  in fact solves the synchronous gauge equations for spatially constant perturbations, this solution cannot be extended to spatially varying perturbations. As argued by Weinberg, the appropriate adiabatic perturbations in  synchronous gauge must come from the solution of the longitudinal gauge equations to next-to-leading order in the long-wavelength expansion. But the latter is in general unknown. 

In synchronous gauge, the conventional approach to determine appropriate adiabatic initial conditions involves an expansion of the linearized solutions in powers of conformal time $\tau$, which one can use to find appropriate initial conditions in the long-wavelength limit $k\tau\to 0$. In order to do so, one has to expand the scale factor and energy densities in powers of $\tau$ \cite{Bucher:1999re}. This does not pose any technical problem in the standard scenario, but it the presence of annihilation in fails because, from equation (\ref{eq:c bkgd evolution}), the dark matter density is non-analytic around $\tau=0$. More generally, an expansion around $\tau=0$ requires assumptions about the evolution of the universe around the time of the big-bang, which is precisely the time around which we know the least about the universe.

In the specific case of coupled  fluids, however, Malik and Wands have shown that  the linearized perturbation equations in any gauge admit an adiabatic solution in the long-wavelength limit with 
\begin{equation}\label{eq:adiabatic}
\frac{\delta\rho_\alpha}{\rho'_\alpha}=
\frac{\delta\rho_\beta}{\rho'_\beta}
\end{equation}
if the intrinsic non-adiabatic energy transfer of each individual fluid $\delta Q_{\mathrm{intr},\alpha}$ vanishes  \cite{Malik:2004tf}.  To check whether this is true in our case, we note that during radiation domination we can neglect the influence of dark matter and dark radiation perturbations on the metric potentials. In that limit,  the adiabatic solution for the dominant constituents takes its conventional form \cite{Bucher:1999re},
\begin{subequations}\label{eq:initial}
\begin{align}
\label{eq:adiabatic densities}
\eta &=-\zeta_i, & h&=-\frac{\zeta_i}{2}(k\tau)^2, &\delta_\gamma &=\frac{\zeta_i}{3}(k\tau)^2, &\delta_\nu&=\delta_\gamma,  &\delta_b &=\frac{3}{4}\delta_\gamma, \\
\label{eq:adiabatic velocities}
& {} & {} & {} &v_\gamma &=-\frac{\tau}{12}\delta_\gamma, 
&v_\nu &= \frac{23+4R_\nu}{15+4R_\nu}v_\gamma, &v_b&=v_\gamma, 
\end{align}
where the normalization has been chosen so that the curvature perturbation  equals $\zeta_i$ (along this adiabatic solution $\zeta$ is conserved),  and we only quote the leading terms in the long-wavelength expansion, since the subleading corrections depend on the unknown behavior of dark matter around $\tau=0$.  Then, it is simple to check using equations (\ref{eq:delta sigma v}), (\ref{eq:adiabatic}) and (\ref{eq:adiabatic densities})  that the  intrinsic energy transfer of dark matter 
\begin{equation}
\delta Q_{\mathrm{intr},c}=- 
	\left(\frac{\delta\langle \sigma v\rangle}{m}-\frac{\langle\sigma v\rangle'}{m}
	\frac{\delta\rho_c}{\rho_c'}\right)\rho_c^2 
\end{equation}
 vanishes, because in the Sommerfeld regime ${\langle\sigma v\rangle'=\mathcal{H}\langle\sigma v\rangle}$. We can therefore specify  initial conditions for dark matter and dark radiation using equation (\ref{eq:adiabatic}),
\begin{equation}\label{eq:delta adiabatic}
\delta_c=\left(\frac{3}{4}+\frac{\langle \sigma v\rangle}{4m}\frac{\rho_c a}{\mathcal{H}}\right)\delta_\gamma, \quad
\delta_d=\left(1-\frac{\langle \sigma v\rangle}{4m}\frac{\rho_c^2 a}{\rho_d\mathcal{H}}\right)\delta_\gamma.
\end{equation}
Again, the magnitude of the impact of annihilation on the initial conditions is determined by the ratio $R$ in equation (\ref{eq:impact}). 

On the other hand, the adiabatic solution discussed in reference \cite{Malik:2004tf} does not constrain the velocity perturbations. In order to determine the latter we note that equation (\ref{eq:vd prime}) has the integral solution
\begin{equation*}\label{eq:rho v d}
	\rho_d v_d=-\frac{1}{4a^4}\int_{\tau_i}^\tau a^4 \delta\rho_d,
\end{equation*}
where we have assumed that at $\tau_i$, $\rho_d v_d$ vanishes. For $\tau_i\ll \tau$ this reproduces for instance the conventional adiabatic solution if we replace dark radiation by standard radiation in the last equation. The dark radiation density perturbation can be found using equation (\ref{eq:delta adiabatic}),  which to first order in $\langle \sigma v\rangle$  gives
\begin{equation}\label{eq:v d ini}
	v_d=v_\gamma+\frac{1}{64}\frac{\langle \sigma v\rangle}{m}
	\frac{\rho_c a}{\mathcal{H}}
	\frac{\rho_c}{\rho_d}\tau \delta_\gamma.
\end{equation}
This expression reduces to the standard adiabatic solution in the limit $\langle \sigma v\rangle\to 0$, even though the second term on the right-hand-side typically dominates when $\rho_d$ is very small.  In our numerical code we use $\rho_d v_d$ as an independent variable, which, according to equation (\ref{eq:v d ini}) has a well-defined value even if $\rho_d$ is initially zero.
\end{subequations}
Recall that $v_c\equiv 0$ by gauge choice.

\subsection{Impact on Structure Formation}

Annihilations impact the growth of the perturbations on many fronts: The evolution of the background differs from the one without annihilations, the evolution of the perturbations differ from their counterparts without annihilation, and, finally, the initial conditions differ from their counterparts in the absence of annihilation.

Because annihilation enters the equations that model dark matter annihilation only through  the combination $\langle \sigma v\rangle/m$, in the limit we are considering, cosmological observables are only sensitive to the combination $\langle \sigma v\rangle/m$.  Here, as throughout this work, we focus on the case of Sommerfeld-enhanced annihilation, in which the average cross section times velocity is proportional to the scale factor. It is thus convenient to introduce an appropriately normalized constant $\Gamma_w$ implicitly determined by
\begin{equation}\label{eq:gamma}
	\frac{\langle \sigma v\rangle}{m c^2}\equiv\Gamma_w 
	\frac{\langle \sigma v\rangle_w}{\mathrm{TeV}}\, a,
\end{equation}
where $\langle \sigma v\rangle_w$ is defined in equation (\ref{eq:sigma0}). Thus, because in our conventions $a$  equals one today, $\Gamma_w$ is the present value of $\langle \sigma v\rangle/mc^2$ in units of $\langle \sigma v\rangle_w/\mathrm{TeV}$.

\begin{figure}[H]
\begin{center}
\subfloat{\includegraphics[width=0.5\textwidth]{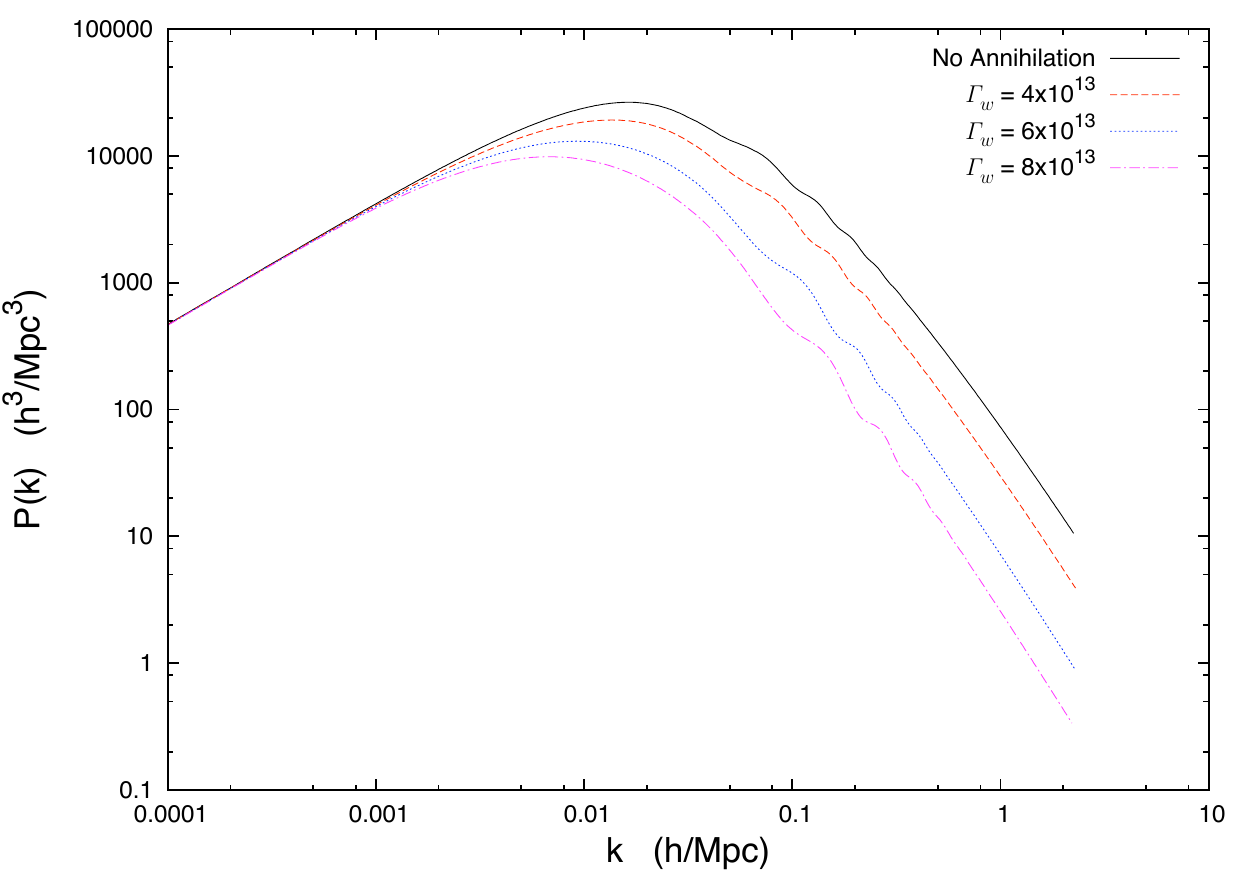}}
 \subfloat{\includegraphics[width=0.5\textwidth]{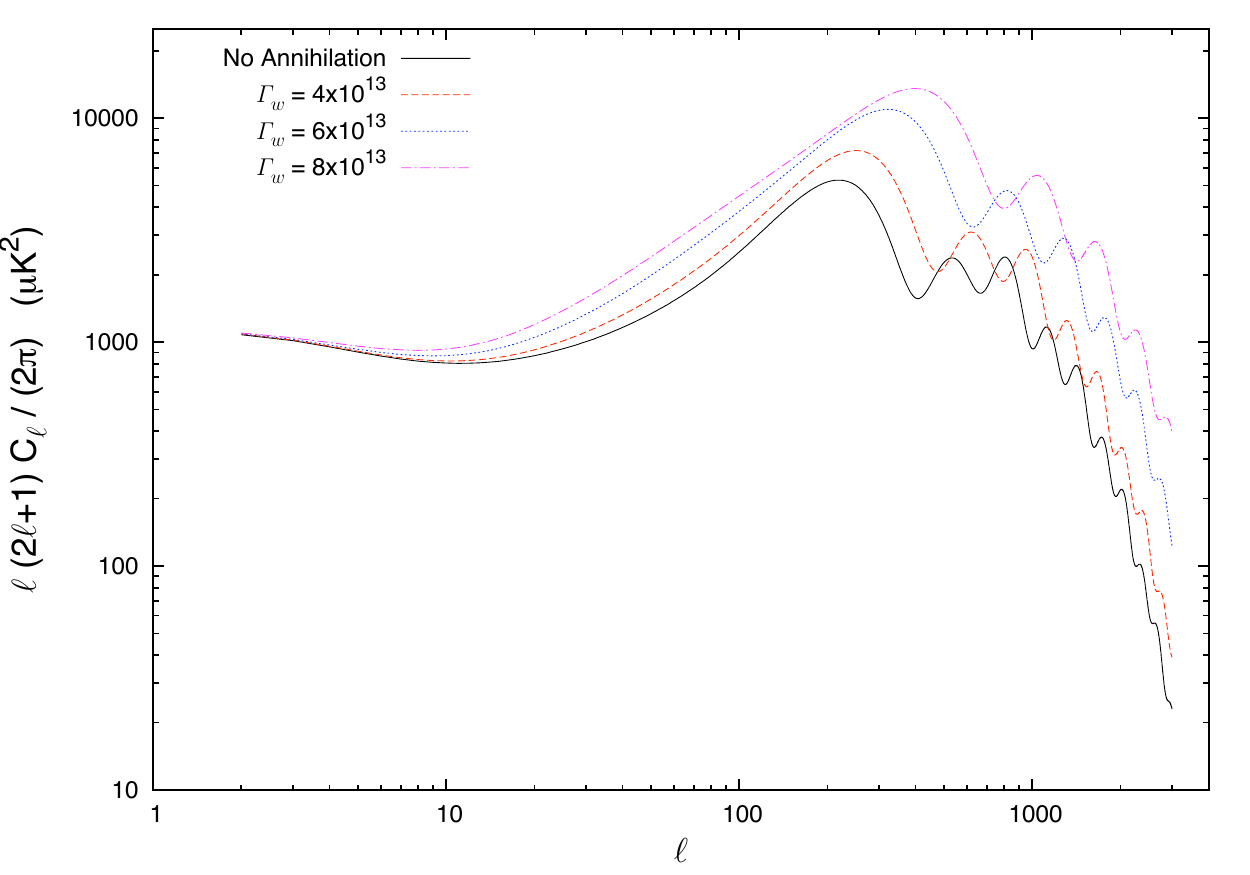}}
\end{center}
\caption{Matter and temperature anisotropy  power spectra for different values of $\Gamma_w$. All the remaining cosmological parameters, including $\Omega_c^0$, are kept fixed.}\label{fig:impact}
\end{figure}

To determine the precise effects of annihilation on the CMB and matter power-spectrum we have modified the Boltzmann integrator CAMB \cite{camb,Lewis:1999bs} by including the three contributions mentioned above. In figure \ref{fig:impact}, we plot the matter and temperature anisotropy  power spectra for different values of $\Gamma_w$, while keeping the remaining cosmological parameters fixed. The impact of annihilation on the CMB power spectrum is visible only for relatively large values of $\Gamma_w$, for which the amount of dark radiation is significant. This dark radiation is what drives most of the impact on the power spectra in this regime. In particular, dark radiation delays the onset of matter-domination, which shifts the wave number of the mode that enters the horizon at matter-radiation equality, $k_\mathrm{eq}$ to larger scales.  On small scales ($k\ll k_\mathrm{eq}$) this shift has no effect, because the transfer function approaches a constant, whereas at larger scales ($k\gg k_\mathrm{eq}$), the power is suppressed by the corresponding factor of $(k_\mathrm{eq}/k_\mathrm{eq}^0)^2$ from the transfer function, where $k_\mathrm{eq}^0$ is the mode that enters at equality in the absence of annihilation. Accordingly, the maximum of the power spectrum at $k=k_\mathrm{eq}$ is shifted to smaller values of $k$, as seen in the left panel of figure \ref{fig:impact}.

The delay in matter-radiation equality  also affects the size of the sound horizon at recombination, which becomes smaller because, with the remaining parameters fixed, the latter is a monotonically growing function of the  redshift at matter-radiation equality (see for instance \cite{Weinberg:2008zzc}.) Hence, the angular size of the sound horizon at recombination decreases, thus shifting the cosmic microwave acoustics peaks to higher values of $\ell$. Apart from Silk damping at very small scales, the amplitudes of these acoustic peaks depend on a monotonically growing function of $k/k_\mathrm{eq}$. Hence, a shift in $k_\mathrm{eq}$ to smaller values causes the anisotropy at a given angular scale (fixed value of $k$) to increase, as observed on the right panel of figure \ref{fig:impact}.

For smaller (and more realistic) values of $\Gamma_w$, the shift in matter-radiation equality is not as pronounced, and an accurate description of the impact of annihilation becomes impractical, because no single effect dominates the phenomenological signatures of annihilation.

\section{Results}

Because annihilation affects both the cosmic microwave temperature anisotropies and the matter power spectrum, measurements of the latter  place constraints on how strongly dark matter annihilates.  We can obtain a rough estimate of the kind of limits that we should be able to impose on $\Gamma_w$, defined in equation (\ref{eq:gamma}), by estimating the Fisher information. As we argued above, the impact of annihilation is dictated by the magnitude of $R$  in equation (\ref{eq:impact}), so  on dimensional grounds we expect $\Delta C_\ell/C_\ell \sim R$.  At leading order, $R$ is constant during radiation domination and decays during matter domination. Replacing $R$ by $R_{\mathrm{eq}}$, and assuming that the temperature multipoles are normally distributed, the Cramer-Rao bound on the variance of an estimate of $R_\mathrm{eq}$ leads to 
\begin{equation}\label{eq:rough limit}
	\Delta\Gamma_w\lesssim \frac{8\pi G}{3c^2}\frac{a_\mathrm{eq}^{1/2}}{H_0}
	\frac{\mathrm{TeV}}{\langle \sigma v\rangle_w} \frac{1}{\sqrt{\ell{}\,^2_\mathrm{max}}}
	\approx \frac{10^{14}}{\ell_\mathrm{max}}.
\end{equation}
Here, $\ell_\mathrm{max}$ is the maximum  multipole probed by the WMAP and ACBAR missions,  ${\ell_\mathrm{max}\approx 10^3}$. As we shall see, the rough estimate  in equation (\ref{eq:rough limit})  is  in fact not far from the actual standard deviation of $\Gamma_w$ that we calculate later.

\begin{figure}
\begin{center}
\subfloat{\includegraphics[width=0.65\textwidth]{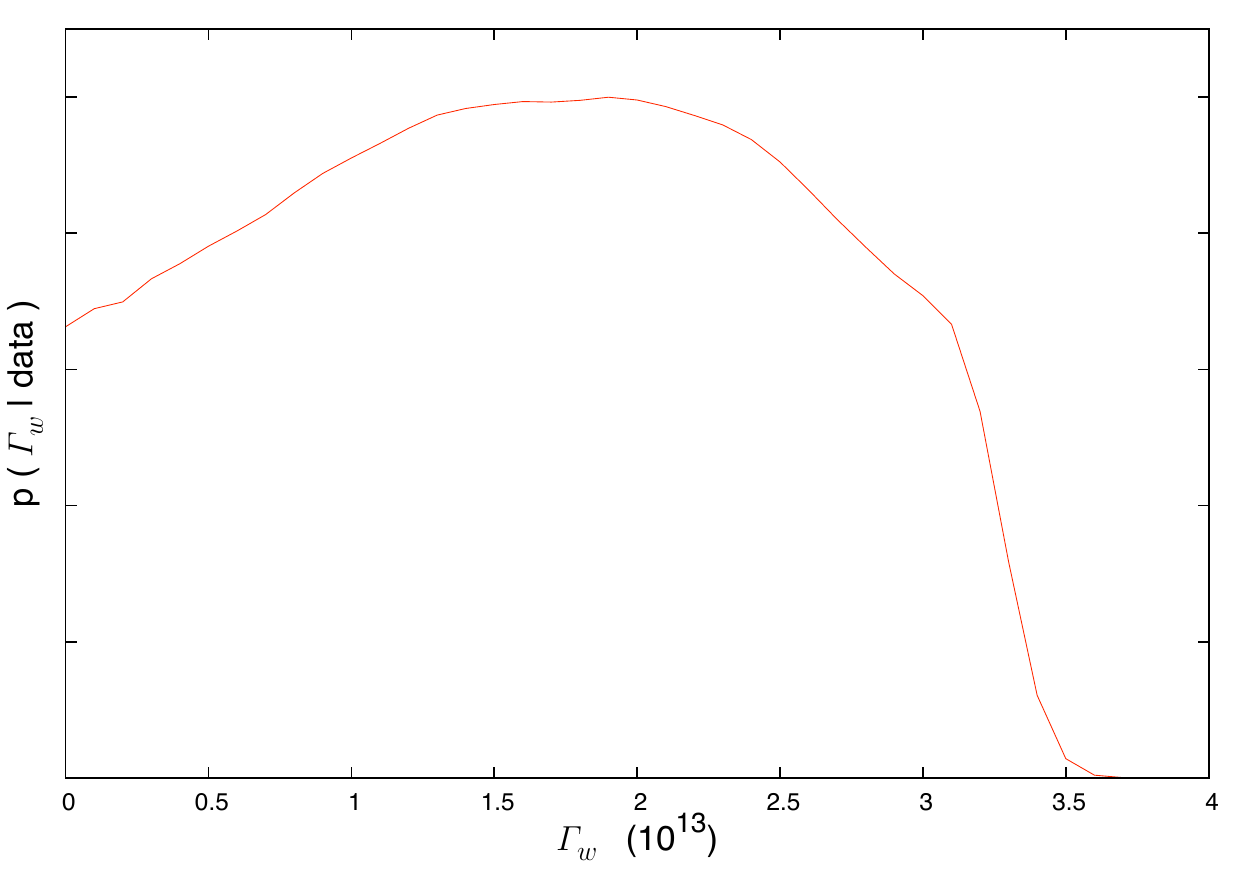}}
\end{center}
\caption{Smoothed marginalized posterior probability distribution function of $\Gamma_w$.}\label{fig:pdf}
\end{figure}

\subsection*{Upper Limits}
To obtain upper limits on the value of $\Gamma_w$ we follow the standard  Bayesian approach in cosmological parameter estimation. We sample the posterior probability for a cosmological model with parameters $H_0$ (Hubble's constant today), $\Omega_\Lambda^0$ (critical density fraction of a cosmological constant), $\Omega_b^0 h^2$ (baryon density), $\tau$ (optical depth), $n_s$ (scalar spectral index), $A_s$ (scalar spectral amplitude) and $\Gamma_w$ in equation (\ref{eq:gamma})  with a set of four Monte Carlo Markov chains of $2.5\times 10^5$ elements each,  generated with an appropriately modified version of COSMOMC \cite{COSMOMC,Lewis:2002ah}. We impose flat priors on all  parameters, assume that the universe is spatially flat and neglect tensor modes. To check for the converge of our chains, we monitor the Gelman and Rubin statistic \cite{Gelman:1992zz}, which stays under $2\times 10^{-3}$ for all parameters. Following COSMOMC output, we also estimate the statistical errors on our upper limits by exploring their changes upon split of our chains in several subsamples; the corresponding relative  errors   remain below  $1\%$.
 
To derive our first limits, we use cosmic microwave temperature anisotropy and polarization data from the seven year WMAP release \cite{Komatsu:2010fb}, small angular scale temperature anisotropy data from the ACBAR experiment \cite{Reichardt:2008ay}, and large scale structure data from an SDSS luminous red galaxy (LRG) sample \cite{Reid:2009xm}.  For the WMAP and LRG data sets, the likelihood of a model is calculated using the codes supplied by the corresponding collaboration.  The (smoothed) marginalized posterior probability density of $\Gamma_w$  is shown in figure \ref{fig:pdf}. The posterior mean and standard deviation of $\Gamma_w$ are
\begin{equation}
	\langle \Gamma_w\rangle=1.65\cdot 10^{13}, \quad \sqrt{\langle \Gamma_w^2\rangle^2-\langle \Gamma_w\rangle^2}=8.94\cdot 10^{12},
\end{equation}
which suggests that there is no significant evidence for dark matter annihilation. In fact, the highest density set\footnote{In our context, a highest density set is a credible interval of prescribed probability content and minimal length. See, for instance, 2.50 in  \cite{Kendall}.} with probability content $p=95\%$ contains $\Gamma_w=0$, which confirms that the latter is in reasonable agreement with the data.

In order to settle whether there is evidence for a non-zero value of $\Gamma_w$, we focus  on the likelihood of the data under the two hypotheses
\begin{equation}\label{eq:hypotheses}
	\left\{
		\begin{array}{l l} 
			H_0: & \Gamma_w=0, \\
			H_1: & \Gamma_w\neq 0. \\
		\end{array}
	\right.
\end{equation}
The Bayes factor (the ratio of  marginalized likelihoods under both hypotheses) is often advocated in Bayesian hypothesis testing. Unfortunately, for nested hypothesis  of the form (\ref{eq:hypotheses}), it is ill-defined for improper priors, or very sensitive to the width of any  uninformative  prior placed on  the additional parameters (see e.g. 7.17 in \cite{Kendall}). We focus instead on the  likelihood ratio
\begin{equation}
	\lambda\equiv \frac{\max_{H_0} L(\textrm{data}|H_0)}{\max_{H_1}  L(\textrm{data}|H_1)},
\end{equation}
which is a statistic that has often proved to be sensible in the classical context, and is closely connected to the Bayes factor asymptotically. Evaluating the maximum likelihoods under both hypotheses, we find
\begin{equation}
	-2 \log \lambda=0.92.
\end{equation}
Recall that under $H_0$, $-2\log \lambda$ is asymptotically distributed like $\chi^2$ with $1$ dof, so the evidence against the null hypothesis $H_0$ is weak at best.\footnote{Both the distribution of $-2\log \lambda $ and its relation to the Bayes factor in the form of the Schwarz information criterion are often derived in the limit of a large number of independent and identically distributed variables. Although the temperature multipoles $a_{\ell m}$ are indeed independent in a statistically isotropic universe, they are however not identically distributed. Hence, care should be  when quoting precise statistical predictions based on likelihood ratios.}  
 
 We thus proceed to set an upper limit on the value $\Gamma_w$. From the posterior distribution we finally derive the  $95\%$ credible upper limit
 \begin{equation}
 \Gamma_w\leq 3.09 \cdot 10^{13}.
 \end{equation}
Our results are summarized in table \ref{table:Gamma}.

\begin{table}
\begin{center}
\begin{tabular}{c c c c c}
Dataset &  $\mu$ & $\sigma$   & $68\%$ & $95\%$ \\
\hline
 WMAP+ACBAR+LRG & $1.65\cdot 10^{13}$ & $8.94\cdot 10^{12}$ 
	& $\,{}\leq 2.18\cdot 10^{13}$ & $\,{}\leq 3.09 \cdot 10^{13}$ \\
\hline
\end{tabular}
\end{center}
\caption{Posterior mean $\mu$ and standard deviation $\sigma$ of $\Gamma_w$, and $68\%$ and $95\%$ upper credible limits on $\Gamma_w$.}
\label{table:Gamma}
\end{table}

In our conventions $a=1$ today, so the previous limits translate for instance into ${\langle \sigma v\rangle/mc^2\lesssim 2.81 \cdot 10^{10}}\langle\sigma v\rangle_w/\mathrm{TeV}$ around recombination, at $z=1100$.  In that respect, for moderately small values of $f$, our constrain on the value of $\langle \sigma v\rangle$ at recombination is orders of magnitude weaker than the limit (\ref{eq:rec limit}) based on recombination alone. Therefore, in those cases it is safe to ignore the impact of annihilation on the evolution of structure. But in any case, our limits are fundamentally different from (\ref{eq:rec limit}) because while the latter only constrains $\langle \sigma v\rangle/m$ at recombination, the former are sensitive to the evolution of  $\langle \sigma v\rangle/m$ throughout cosmic history. Since WMAP is sensitive to comoving scales with $k\, \tau_0\sim 10^3$, our limits are sensitive to the value of $\langle \sigma v\rangle$ at redshifts of about $z\sim 10^4$. 
  
The remaining cosmological parameters ($H_0, \Omega_{\Lambda}^{0}, \Omega_b^0 h^2, \tau, n_s, A_s$) do not differ significantly from their values with $\Gamma_w = 0$. In particular, their best fit values under $H_1$ fall within the posterior $95\%$ credible limits on the corresponding parameters under $H_0$. The parameter $\Gamma_w$ shows the strongest correlations  with the amount of baryons $\Omega_b h^2$ ($-80\%$), the Hubble parameter $H_0$ ($76\%$) and the age of the universe ($-89\%$), although the latter still remain well constrained by the data. The negative correlation between  $\Gamma_w$ and the age of the universe  is, for instance, what we expect from our analysis of the background evolution at the end of subsection \ref{sec:Background Evolution}. 

\section{Summary and Conclusions}

We have studied the impact of dark matter annihilation on the cosmic microwave background and the matter power spectrum, under the assumption that  dark matter annihilates into dark radiation with an averaged cross section times velocity $\langle \sigma v\rangle$ that grows in proportion to the scale factor. This Sommerfeld enhancement is expected to occur generically in any dark matter model in which dark matter particles experience an additional attractive interaction, regardless of the dark matter annihilation channels. Most previous analyses of this scenario assumed that dark matter predominantly annihilates into standard model particles (visible radiation). Our analysis focuses on the purely gravitational impact of the annihilation, and thus holds for a much wider class of models. In particular, within the context of our analysis we can address whether the relativistic dark matter annihilation products consitute the dark radiation that some analyses of cosmic data seem to favor. 

Actual cosmic microwave anisotropy and large scale structure data do not  show evidence of dark matter annihilation with such a growing cross section, so we have derived the  limits on the corresponding  averaged cross section times velocity listed in table \ref{table:Gamma} (we define $\Gamma_w$  in equation (\ref{eq:gamma}).) As seen in the table, these upper limits  allow $\langle \sigma v\rangle$ to be several orders of magnitude larger than a typical weak annihilation cross section. In particular, our limits indicate that if dark matter annihilation deposits a significant fraction of the annihilation energy  into visible radiation, $f\gg 10^{-8}$, the effects on the cosmic microwave background that we have studied here are subdominant. On the other hand,  if $f\ll10^{-8}$ the impact of annihilation on recombination is subdominant, and the gravitational effects that we have studied here play the dominant role.  Because the data do not seem to support the dark matter annihilation hypothesis, we do  not find evidence supporting  an additional dark relativistic species originating from such annihilations either. 

At present, the nature of dark matter  remains a mystery. The limits that we have derived are not only useful in further constraining the properties of dark matter itself, but also in constraining its interactions with other elements of the  sector where it resides. Because large scale structure still  allows for very large annihilation cross sections, the dark sector may in principle  host a dark matter candidate with properties far different from the standard collisionless wimp.

\begin{acknowledgments}
The work of CAP and JTN was supported in part by the NSF under  grant PHY-0855523.
\end{acknowledgments}

\appendix

\section{Microscopic Description}
\label{sec:Boltzmann Equation}

In order to determine the impact of annihilation on the dark matter density we begin with a microscopic description of annihilation. Let us consider the phase space distribution of dark matter particles  in the universe, $f$.  It is useful to resort to a  formulation in which the distribution function depends on the space-time coordinates $\tau$ and  $x^i$, and  on covariant spatial momenta $p_j$, ${f=f(\tau, x^i, p_j)}$ (for simplicity we assume that dark matter particles are spinless.) In that case, the distribution function $f$ is a scalar under diffeomorphisms, and it obeys the Boltzmann equation \cite{Boltzmann}
\begin{equation}\label{eq:Boltzmann NC}
	\frac{p^0}{m}\left[\frac{\partial f}{\partial \tau}
	+\frac{\partial f}{\partial x^i}\frac{dx^i}{d\tau}
	+\frac{\partial f}{\partial p_i}\frac{dp_i}{d\tau}\right]=C[f,f],
\end{equation}
where $m$ is the wimp mass and $dp_i/d\tau$ is dictated by the geodesic equation
\begin{equation}\label{eq:geodesic}
	p^0\frac{dp_j}{d\tau}=\frac{1}{2m}
	\frac{\partial g_{\alpha\beta}}{\partial x^j}p^\alpha p^\beta.
\end{equation}
In the above, $p^0$ should be expressed in terms of the covariant momenta $p_i$ and the spacetime metric. 

Although this is not manifest, the left hand side of equation (\ref{eq:Boltzmann NC} ) is a diffeomorphism scalar. Hence, the collision term $C$ is  a  scalar too, and describes the changes in the distribution function caused by collisions and annihilations.  For definiteness, let us assume that the only relevant processes involve the  annihilation of two  dark matter particles $\chi$  into two spinless particles $\phi$ of four-momenta $q_1$ and $q_2$. Then, the collision term is
\begin{equation}
	C[f,f]=-\frac{(2\pi)^4}{m^2 m_\phi^2}
	\int d_*^4 p_2 \, d_*^4 q_1 \, d_*^4 q_2 \,
	f(x^\mu,p_{1\nu}) f(x^\mu,p_{2\nu})
	R_\mathrm{ann}\, \sqrt{-g}\, \delta^4(p_1+p_2-q_1-q_2),
\end{equation}
where we identify $p_1\equiv p$, and all four-momenta are covariant (as opposed to contravariant.) Note the minus sign in front of the last equation, which reflects that we are  considering annihilation processes only.  

The combination 
\begin{equation}
	d^4_* p\equiv \frac{d^4 p }{\sqrt{-g}}
	\, 2m\, \theta(p_0) \delta(p^2+m^2)=
	\frac{m}{\sqrt{-g}}\frac{d^3p}{p^0}
\end{equation}
is a scalar under diffeomorphism, so $R_\textrm{ann}$ has to be a scalar too. We can thus calculate $R_\mathrm{ann}$ using the standard rules of quantum field theory in a local Lorentz frame, in which 
\begin{equation}
R_\textrm{ann}\equiv p_1^0 p_2^0 q_1^0 q_2^0 |\mathcal{M}_\mathrm{ann}|^2,
\end{equation}
 and $\mathcal{M}_\mathrm{ann}$ determines the $\mathcal{S}$-matrix, $\mathcal{S}=-2\pi i \mathcal{M} \delta^4(p_1+p_2-q_1-q_2)$. Say, for an interaction of the form
\begin{equation}\label{eq:simple model}
	S_\mathrm{int}=\int d^4x \sqrt{-g} \, \frac{\lambda}{4} \chi^2 \phi^2, 
\end{equation} 
where $\chi$ represents the dark matter field and  $\phi$ its (relativistic) annihilation products, ${R_\mathrm{ann}=\lambda^2/16}$ at tree level (we follow the conventions of \cite{Weinberg:1995mt}.) 

In this article, however, we are interested in annihilation processes for  which the annihilation rate is boosted by a factor $S$ from Sommerfeld enhancement,
\begin{equation}
	R_\textrm{ann}=\frac{\lambda^2}{16}\times
S(v_0/v_\mathrm{rel}),
\end{equation}
where $\lambda$ is a constant (not necessarily related to the simple model in equation (\ref{eq:simple model})), and $v_0$ is a constant with dimensions of velocity and $v_\mathrm{rel}$ is the appropriate relativistic expression for the relative velocity \cite{Yoon:1999jd}  
\begin{equation}
v_\mathrm{rel}=
	\frac{\sqrt{-(p_1+p_2)^4-4(p_1+p_2)^2 m^2}}{-(p_1+p_2)^2-2m^2}.
\end{equation}
(Because we are interested in the non-relativistic limit, any diffeomorphism scalar $v_\mathrm{rel}$ that reduces to $|\vec{v}_1-\vec{v}_2|$ at non-relativistic momenta in a local Lorentz frame would suffice).   The factor $S$ describes the enhancement of the cross section. In models in which such an enhancement is caused by an attractive interaction mediated by a light force carrier of mass $m_Y$ coupling to dark matter with amplitude $\lambda_Y$ it  has the form \cite{ArkaniHamed:2008qn}
\begin{equation}
	S(x)\approx 
	\left\{
	\begin{array}{ll}
	\displaystyle{\frac{m}{m_Y \alpha}}, & 
	\displaystyle{\frac{v_\mathrm{rel}}{c} \ll  \frac{2 m_Y}{m}}\\
	 \displaystyle{\frac{v_0}{v_\mathrm{rel}}}, & 
	 	\displaystyle{\frac{2 m_Y}{m}\ll \frac{v_\mathrm{rel}}{c}\ll  2\pi \alpha} \\
	 \displaystyle{1}, &
	 \displaystyle{ 2\pi \alpha\ll\frac{v_\mathrm{rel}}{c}}
	 \end{array}
	 \right., 
\end{equation}
where $\alpha=\lambda_Y^2/(4\pi)$ and $v_0=2\pi \alpha$. For the rest of our analysis we restrict ourselves to the intermediate  regime, in which the enhancement is inversely proportional to the relative velocity,
\begin{equation}\label{eq:S enhancement}
S\approx 
	\frac{v_0}{v_\mathrm{rel}} \quad \quad
	 \left(
	 \frac{2 m_Y}{m}\ll \frac{v_\mathrm{rel}}{c}\ll  2\pi \alpha
	 \right).
\end{equation}
Clearly, this range of velocities is strongly model-dependent, although typically, for light force carriers and not too weak couplings  it  can span several orders of magnitude in $v_\mathrm{rel}$. 

\subsection{Perfect Fluid Description}

The energy momentum tensor of the ensemble of particles described by $f$ is
\begin{equation}\label{eq:EMT}
T_{\mu\nu}= \int d^4_* p \, \frac{ p_\mu p_\nu}{m} f,
\end{equation}
which clearly transforms like a tensor. In order to determine whether this energy momentum is conserved, it is convenient to consider a local inertial frame, in which $g_{\mu\nu}=\eta_{\mu\nu}$ and $\Gamma^\mu{}_{\nu\rho}=0$. Then, using  the Boltzmann equation (\ref{eq:Boltzmann NC}) and general covariance it is easy to show that in an arbitrary coordinate system the energy momentum tensor satisfies
\begin{equation}\label{eq:EMT conservation}
	\nabla_\mu T^{\mu\nu}=\int d^4_* p \, p^\nu \,  C[f,f],
\end{equation} 
since the latter holds in any local inertial frame. Thus, in the absence of annihilation the energy momentum tensor is covariantly conserved, as it should. 

In order to relate the kinetic  to the fluid description, following Eckart \cite{Eckart:1940te},  we define the four-velocity of the fluid to be proportional to the averaged particle velocity,
\begin{equation}\label{eq:four velocity}
	u^\mu \equiv \frac{\langle p^\mu \rangle}{\sqrt{-\langle p^\nu \rangle\langle p_\nu \rangle}},
\end{equation}
where the average of any function $g$ of momentum is defined by
\begin{equation}
\langle g(\vec{p})\rangle =\frac{1}{n} \int d^4_* p \, g(\vec{p}) f,
\end{equation}
and $n$ is the (scalar) particle number density,
\begin{equation}
	n \equiv \int d^4_* p \, f.
\end{equation}
It is simple to show that for \emph{any} distribution the Boltzmann equation (\ref{eq:Boltzmann NC}) implies that in the absence of annihilations the current $n\langle p^\mu\rangle$ is covariantly conserved,
\begin{equation}
\frac{1}{m} \nabla_\mu [n \, \langle p\rangle^\mu ]=-n^2 \langle \sigma v\rangle,
\end{equation}
where we have defined the averaged annihilation cross section times relative velocity,\footnote{Recall that  cross sections are rates per flux,  and that the flux is proportional to the relative velocity between the annihilating particles.}
\begin{equation}\label{eq:sigma v}
	\langle \sigma v\rangle=-\frac{1}{n^2}\int d^4_* p \, C[f,f],
\end{equation}
which of course vanishes in the absence of annihilations. In that case,  the four velocity (\ref{eq:four velocity}) is proportional to the current that captures the conservation of matter. 

We shall assume that the distribution $f$ is such that the energy momentum tensor is well approximated by that of a perfect fluid,
\begin{equation}\label{eq:EMT pf}
	T_{\mu\nu}=(\rho+p)u_\mu u_\nu+ p \,  g_{\mu\nu}. 
\end{equation}
 Using equations (\ref{eq:EMT}) and  (\ref{eq:four velocity}), the energy density thus becomes
\begin{equation}\label{eq:rho}
\rho\equiv T_{\mu\nu} u^\mu u^\nu =-\frac{n}{m}
\frac{\langle p_\mu p_\nu\rangle \langle p^\mu \rangle\langle p^\nu\rangle}
{\langle p_\rho \rangle\langle p^\rho\rangle}.
\end{equation}
In order to determine the pressure we note that equation (\ref{eq:EMT}) implies that $T^\mu{}_\mu=-m\, n$, whereas equation (\ref{eq:EMT pf}) implies that $T^\mu{}_\mu=3p-\rho$.  Therefore, the pressure of the fluid simply is
\begin{equation}
p=\frac{\rho - m n}{3},
\end{equation}
 which clearly shows that only relativistic components, those for which the energy density $\rho$ is larger than the ``rest'' energy $m \,n$,  contribute to the pressure.

\subsubsection*{Pressureless fluids}
By definition, the pressure of a non-relativistic fluid of particles vanishes, which implies that $\rho= mn$, as expected. Looking back at equation (\ref{eq:rho}) and noting that $\langle p_\mu p^\mu\rangle=-m^2$ we see that this is the case if the covariance of the four momentum vanishes,
\begin{equation}
\langle p_\mu p_\nu\rangle=\langle p_\mu \rangle\langle p_\nu\rangle.
\end{equation}
It then follows, using (\ref{eq:four velocity}), that
\begin{equation}\label{eq:u pressureless}
u^\mu =\frac{\langle p^\mu \rangle}{m}.
\end{equation}
Given that the covariance of the momenta vanishes by assumption, it is  natural to assume that the we can also replace the momentum on the rhs of equation (\ref{eq:EMT conservation}) by its average. Then, the conservation equation becomes
\begin{equation}\label{eq:conservation}
\nabla_\mu T^{\mu\nu}=-\frac{\langle \sigma v\rangle}{m} \rho^2  u^\nu.
\end{equation}
Since for a pressureless fluid $\rho$ is proportional to the number density $n$,  equation (\ref{eq:conservation}) also expresses conservation of particle number, as can be seeing by looking at the projection of that equation onto $u_\nu$.  
To conclude, we note that  because  $\int d^4_* q_1  d^4_* q_2 \sqrt{-g}\,  \delta^{(4)}(p_1+p_2-q_1-q_2)$ is a scalar,  in the non-relativistic limit the averaged annihilation rate in a universe with metric (\ref{eq:Friedmann perturbed}) becomes 
\begin{equation}\label{eq:sigma v NR}
\langle \sigma v\rangle \approx \frac{(2\pi)^5}{m^2}\frac{\lambda^2}{16}
\frac{a^2}{(-g)}\frac{1}{n^2}
\int d^3 p_1 d^3 p_2 \, \frac{v_0}{v_\mathrm{rel}}f(\vec{p}_1)f(\vec{p}_2),
\end{equation}
where we have assumed that the relative velocities are in the regime in which  Sommerfeld enhancement is effective,  equation (\ref{eq:S enhancement}), and that all the annihilation products are highly relativistic, $m\gg m_\phi$.

\subsection{Background}
Let us turn our attention now to the evolution of $\langle \sigma v\rangle$ in the unperturbed universe (\ref{eq:unperturbed FRW}). Because of homogeneity and isotropy, the distribution function $f$ can only depend on the magnitude of the momentum $f=f(\tau,p)$,  where
\begin{equation}\label{eq:p}
	p\equiv a\sqrt{g^{ij} p_i p_j}.
\end{equation}
In the wimp scenario dark matter decoupled while being non-relativistic, so it would be natural to  consider a Maxwell-Boltzmann ansatz for the distribution function, but this is problematic because it can be shown, that annihilation does not  preserve this form of the distribution function \cite{Feng:2010zp}.
 
We can nevertheless proceed without making any assumptions about the form of $f$ when the  coupling $\lambda$ is sufficiently small. Namely, because $\langle \sigma v\rangle$  in equation (\ref{eq:sigma v NR}) is already of order $\lambda^2$, to leading order  we can calculate $\langle \sigma v\rangle$ by substituting into (\ref{eq:sigma v NR})  the solution of the Boltzmann equation (\ref{eq:Boltzmann NC}) to zeroth order in $\lambda$,
\begin{equation}\label{eq:Boltzmann background}
	\frac{\partial f}{\partial \tau}=0.
\end{equation}
In this case, \emph{any} distribution function $f=f(p)$ solves  equation (\ref{eq:Boltzmann background}), and the density of dark matter particles (to zeroth order)  evolves as we would expect in the absence of annihilation,
\begin{equation}
	n=\frac{1}{a^3}\int d^3 p\, f(p),
\end{equation}
provided that $f$ has support for  non-relativistic momenta only.  Using this form of the density and equation (\ref{eq:sigma v NR}) the averaged cross section becomes
\begin{equation}\label{eq:SE sigma v}
	\langle \sigma v\rangle= \frac{(2\pi)^5}{m^2}\frac{\lambda^2}{16}
	\frac{a\int d^3 p_1 d^3 p_2 \,  f(p_1) f(p_2) \, m  v_0/|\vec{p}_1-\vec{p}_2|}{\left(\int d^3 p f(p)\right)^2}.
\end{equation}
The crucial point is that the the thermal average is proportional to the scale factor $a$, simply because the relative velocity $v_\mathrm{rel}$ between dark matter particles redshifts as the universe expands. 
  
\subsection{Perturbations}

In a perturbed universe (\ref{eq:Friedmann perturbed}) we also need to consider the perturbations in the annihilation cross section, $\delta\langle \sigma v\rangle$.    Again, in the non-relativistic limit it is possible to do so for an arbitrary background distribution and arbitrary metric perturbations by focusing on the leading  result in a small-coupling expansion. In particular, we can calculate $\delta\langle \sigma v\rangle$  to leading order in $\lambda$ by solving the perturbed Boltzmann equation for $\delta f$ to zeroth order   and substituting the corresponding solution into equation (\ref{eq:sigma v NR}).  In doing so, we shall be able to remain in the perfect fluid approximation, without the need to include the evolution of $\delta f$ into our system of perfect fluid equations.

Following \cite{Weinberg:2006hh} let us write the perturbed distribution function as
\begin{equation}\label{eq:decomposition}
f(\tau, \vec{x}, \vec{p})=\bar{f}(p)+\delta f(\tau, \vec{x}, \vec{p}),
\end{equation}
where $\bar{f}$ is an arbitrary distribution with support at non-relativistic momenta,  and $p$ is the magnitude of the spatial momentum defined in (\ref{eq:p}),
\begin{equation}
	p=\sqrt{ p_k p_k}-\frac{1}{2}\frac{h_{ij} p_i p_j}{\sqrt{p_k p_k}}.
\end{equation}
Here and in the following,  Einstein's summation convention is implied even if repeated indices are not in opposite locations.  Because $p$ now depends  on the metric,  $\bar{f}$ also contributes to the perturbations of the distribution function. Then, the perturbation $\delta f$ obeys the linearized Boltzmann equation 
\begin{equation}\label{eq:linearized Boltzmann}
	\frac{\partial \delta f}{\partial\tau}+\frac{\partial \delta f}{\partial x^i}\frac{1}{a^2}\frac{p_i}{p^0}-\frac{1}{2}\frac{\bar{f}'}{p} 
	\frac{\partial h_{jk}}{\partial \tau} p_j p_k=0,
\end{equation}
where a prime denotes derivative with respect to the argument ($p$ in this case). Recall that we set the collision term to zero because we are only interested in evaluating $\delta \langle \sigma v\rangle$ to zeroth order in $\lambda.$

The linearized Boltzmann equation (\ref{eq:linearized Boltzmann}) has the line of sight solution
\begin{equation}
	\delta f=\frac{1}{2}\frac{\bar{f}'}{p}\int_{\tau_i}^\tau d\tau' p_i p_j \,
	h_{ij,\tau}\!\left(\tau', \vec{x}-\int_{\tau'}^\tau d\tau'' \frac{1}{a}\frac{\vec{p}}{m}\right),
\end{equation}
which assumes that $\delta f$ was negligible at the initial time $\tau_i$, and that dark matter is non-relativistic. In general, this solution is a non-local functional of $h_{ij}$, but in the non-relativistic limit in which $p/m\ll 1$, we can set the momentum in the argument of the integral to zero, which yields a simple local expression for $\delta f$ in terms of the metric perturbations $h_{ij}$,
\begin{equation}\label{eq:delta f}
	\delta f(\tau,\vec{x},\vec{p})=\frac{1}{2}\frac{\bar{f}'}{p} h_{ij}(\tau,\vec{x})\,  p_i p_j,
\end{equation}
where we have assumed again that the  perturbations of $h$ are initially negligible (as we discuss in subsection \ref{sec:Initial Conditions}, this holds  for adiabatic initial conditions.) Note that the first correction to this result away from the strict non-relativistic limit would be proportional to three momenta, and would therefore vanish in momentum integrals invariant under rotations like the ones involved in the calculation of $\delta \langle \sigma v\rangle$. Given the structure of the terms we have omitted, we  expect this approximation to be valid on scales
\begin{equation}
 (k \tau)^2\ll \left(\frac{m}{p/a}\right)^2,
\end{equation}
which for non-relativistic momenta encompasses modes well within  the horizon. Since momenta redshift with $a$, this approximation becomes increasingly accurate. 

With the  explicit expression for $\delta f$ in equation (\ref{eq:delta f}) at hand, we can calculate $\delta \langle \sigma v \rangle$ by substituting the solution (\ref{eq:delta f}) into equation (\ref{eq:SE sigma v}). Because the latter is a function of $g_{ij}$, invariant under spatial diffeomorphisms, metric perturbations do not contribute to $\delta\langle \sigma v\rangle$, and we may restrict our attention directly to the contributions from $\delta f$ alone. The resulting integrals can be simplified by noting that rotational invariance and linearity demand that $\delta \langle \sigma v \rangle $ be proportional to the trace of $h_{ij}$, and explicit calculation shows that 
\begin{equation}\label{eq:delta sigmav SE}
\delta \langle \sigma v\rangle=\frac{h}{6}\langle\sigma v\rangle.
\end{equation}
In this way, the system of perfect fluid equations remains closed, and there is no need to track the evolution  of $\delta f$ in our system of equations. Also note that the velocity perturbation associated with the solution (\ref{eq:delta f}) vanishes, and is therefore consistent with the gauge choice $v_c=0$ in equation (\ref{eq:vc zero}).

It is also instructive to explore how $\delta f$ is affected by gauge transformations, and how the residual gauge symmetry allowed by synchronous gauge leads to the existence of gauge mode solutions. Under a gauge transformation
\begin{equation}
	x^\mu \to \tilde{x}^\mu=x^\mu+ \epsilon^\mu
\end{equation}
the perturbation in the distribution function $\delta f$ defined in equation (\ref{eq:decomposition}) transforms as
\begin{equation}\label{eq:delta f gauge}
	\Delta \delta f\equiv \delta\tilde{f}-\delta f=
	\frac{\bar{f}'}{p} \left( \epsilon_{,ij} \,  p_i p_j+\epsilon^0{}_{,i}\,  p_i p_0\right)+
	\frac{1}{2}\frac{\bar{f}'}{p} p_i p_j \Delta h_{ij},  
\end{equation}
where we have used that $\epsilon^i\equiv \partial_i \epsilon$ in the scalar sector. 
The additional term proportional to $\Delta h_{ij}$ originates from the dependence of $\bar{f}$ on the metric perturbations. Under the same gauge transformations, the latter transform as
\begin{subequations}\label{eq:h gauge}
\begin{align}
	\Delta h_{00}&=2\mathcal{H} \epsilon^0 +2 \epsilon^0{}_{,\tau} 
	\label{eq:delta h 00} \\
	\Delta h_{0i}&=\epsilon^0{}_{,i}-\epsilon_{,i\tau} 
	\label{eq:delta h 0i} \\
	\Delta h_{ij} &=-2\epsilon_{,ij}-2\mathcal{H} \epsilon^0 \delta_{ij}.
\end{align}
\end{subequations}
Equations (\ref{eq:delta h 00}) and (\ref{eq:delta h 0i}) immediately reveal that synchronous gauge contains a residual gauge freedom. A coordinate transformation with 
\begin{equation}\label{eq:epsilon}
	\epsilon^0=\frac{A(x)}{a}, \quad \epsilon=B(x)+A(x)\int_{}^\tau \frac{d\tau'}{a(\tau')}
\end{equation}
preserves the synchronous conditions $h_{00}=h_{0i}=0$, and thus leads to the existence of gauge modes. In fact, it is easy to check that equations (\ref{eq:delta f gauge}) and (\ref{eq:h gauge}), with $\epsilon^\mu$ given by equations (\ref{eq:epsilon})  solve the linearized Boltzmann equation (\ref{eq:linearized Boltzmann}). Substituting this gauge mode into expression (\ref{eq:sigma v NR}) we find that the term proportional to $p_i p_0$ does not contribute to $\delta\langle \sigma v\rangle$ because of rotational invariance. In the remaining terms, the factors of $\epsilon$ cancel,  so the corresponding $\delta \langle \sigma v\rangle$ equals what we would get  from  a  perturbation $\delta f$  of the form (\ref{eq:delta f}) with an effective metric perturbation 
\begin{equation}
	h_{ij}=-2\mathcal{H} \epsilon^0 \delta_{ij}. 
\end{equation}
If we substitute this effective metric perturbation into equation (\ref{eq:delta sigmav SE}) we find that
 \begin{equation}
 	\delta \langle \sigma v\rangle=-\mathcal{H} \langle \sigma v\rangle \epsilon^0.
 \end{equation}
This is precisely what we expect from a gauge transformation of a scalar proportional to the scale factor, and it also leads to a (gauge mode) solution of the perturbed equations (\ref{eq:delta c prime}).   This agreement thus provides a check of expression (\ref{eq:delta sigmav SE}) and the consistency of our approach.

\end{document}